# Real Space Imaging of Spin Scattering in Chirality-Induced Spin Selectivity

*Jaehyun Lee[1]‡, Sang-hyuk Lee[1]‡, Uiseok Jeong[1], Daryll J. C. Dalayoan[1], Soobeom Shin[2], Hu Young Jeong[2], Hosub Jin[1], Binghai Yan[3,4]\*, Noejung Park[1, 5]\*, Seon Namgung[1]\**

[1]Department of Physics, Ulsan National Institute of Science and Technology (UNIST), Ulsan, Republic of Korea

[2]Graduate School of Semiconductor Materials and Devices Engineering, Ulsan National Institute of Science and Technology (UNIST), Ulsan, Republic of Korea

[3]Department of Condensed Matter Physics, Weizmann Institute of Science, Rehovot, Israel

[4]Department of Physics, Pennsylvania State University, PA, US

[5]Max Planck Institute for the Structure and Dynamics of Matter, Hamburg, Germany

*Correspondence to: seon@unist.ac.kr, noejung@unist.ac.kr, binghai.yan@weizmann.ac.il

KEYWORDS: chirality-induced spin selectivity, tellurium, magneto-optics, spintronics

The interaction between electron spin and molecular chirality plays a fundamental role in quantum phenomena, with significant implications for spintronics and quantum computing. The chirality-induced spin selectivity (CISS) effect, where chiral materials preferentially transmit electrons of a particular spin, has sparked intense interest and debate regarding its underlying mechanism. Despite extensive research, the spatial distribution of spin polarization in chiral systems, the key evidence to reveal the spin scattering mechanism in CISS, has remained experimentally elusive particularly due to complications arising from spin-orbit coupling in metal electrodes typically used in such studies. Here we show, through reflective magnetic circular dichroism measurements on chiral tellurium nanowires with graphene electrodes, that current-induced spin polarization exhibits identical signs in both the nanowire and electrodes, distinct from the presumed spin filter scenario. The observed spin polarization scales linearly with current amplitude, aligns parallel to the current direction, reverses with chirality or current flow, and demonstrates spin relaxation lengths of several micrometers into graphene. Our findings provide the first direct visualization of spatial spin distribution in chiral devices. This work establishes a new paradigm for investigating spin-dependent phenomena in chiral materials and opens avenues for developing chirality-based spintronic and quantum devices.

INTRODUCTION

Chirality, a fundamental property of matter, has emerged as a powerful tool for manipulating electron spin in nanoscale systems and sparked intense research interest due to its potential applications in spintronics and spin-based quantum computing[1-6]. Understanding and harnessing spin polarization in chiral systems is essential to unlocking novel quantum phenomena and developing next-generation spintronic devices but remains elusive and debated[7,8]. In CISS, a spin filter scenario[1] is widely presumed where current flow or spontaneous charge redistribution (referred to as electron displacement current in literature[9,10]) leads to opposite spin polarization at two terminals of the chiral material (Figure 1). This mechanism was frequently adopted to interpret CISS-driven phenomena, such as the giant magnetoresistance[9,10], anomalous Hall effect[11-13], selective enantiomer adsorption on a magnetic substrate[13-17], and enhanced electrocatalytic activity via spin polarization in chiral systems[18]. However, the spin filter was pointed out to violate some basic principles[19-21], for example, it can lead to unphysical equilibrium spin current[22]. Instead, a spin polarizer mechanism[19] was predicted where transmitted and reflected electrons exhibit the same spin polarization (Figure 1). Upon reversing the current or electron displacement, the spin filter predicts unchanged spin polarization at a given terminal while the spin polarizer predicts flipped spin polarization. A recent magnetoresistance measurement on chiral organic conductors[23] claimed to support the spin filter scenario, while a more recent experiment favored the spin-polarizer model by studying spin current injection into the chiral materials[24]. Therefore, direct real space spin imaging on a well-controlled device is urgently demanded to unambiguously resolve the essential spin polarization of CISS.

In this work, we detect the nonequilibrium spin polarization using reflective magnetic circular dichroism (RMCD) measurement of current-flowing chiral tellurium (Te) nanowire between two graphene electrodes. In previous experimental studies, organic chiral molecules were typically measured in contact with heavy metal electrodes or substrates, in the form of a vertical junction. The origin of spin-orbit coupling (SOC) in these systems—whether it stems from proximity to heavy metals[25-27] or is intrinsic to the organic molecules—has been a subject of intense debate[7], hindering our understanding of spin distribution. To overcome this challenge, in the present work, we deliberately designed a system of interest in a reverse manner: the chiral nanowire with strong intrinsic SOC in contact with electrodes that have negligible SOC, corresponding to the Te nanowire and graphene, respectively. This strategy would provide clear insights into the CISS process. By scanning RMCD on planar devices, we can observe the spin polarization of each component, the chiral nanowire and both electrodes, which was difficult or impossible in previous vertical molecular junctions. We find that the current-induced spin polarization exhibits the same sign in the nanowire and both electrodes and depends linearly on the current, and its sign reverses upon flipping chirality or current direction. This work provides the first optical visualization of the CISS effect, reveals a spin polarizer mechanism, and opens new avenues for designing chirality-based spintronic and optoelectronic devices.

RESULTS

**Crystal and band structure of Te.** Te is an elemental semiconductor with nonmagnetic chiral atomic chains bonded via van der Waals forces. Attributed to the chirality in the non-centrosymmetric structure of Te, noteworthy observations have been reported including nonreciprocal transport in magnetic field[28-30], circular photo-galvanic effect (CPGE)[31,32], topological signatures[34-36], and current-induced magnetization[37-40]. The chiral crystalline structure of a Te nanowire is shown Figure 2a. The unit cell contains three atoms that are connected through a threefold screw rotation, which can be either right- or left-handed along the c-axis. In our study, single crystalline Te nanowires were synthesized by a hydrothermal method[28,41], and the synthesized Te nanowires have dimensions of tens of micrometers in length and hundreds of nanometers in width (Figure 2b). A scanning transmission electron microscopy (STEM) image of a Te nanowire along the c-axis shows covalent bonds of Te helical atomic chains arranged in triangles (Figure 2c). Figure 2d shows STEM images of Te nanowires with two irreducible images classified as enantiomorphic chiral space groups $P3_121$ or $P3_221$, corresponding to right- and left-handed Te nanowires, respectively[42,43]. The well-defined two chirality of nonmagnetic Te nanowires with strong SOC make it well-suited for studying CISS. The energy band structures with the projected spin ($S_z$) of each chiral wire are presented in Figure 2e, which were calculated with the ab-initio density functional theory (DFT). The detailed computational methodology is provided in the Methods section. Under the pronounced SOC of Te, the bands clearly exhibit spin–momentum-locked textures. Notably, the spin directions of the two chiral wires along the Γ–A path are opposite as indicated by the arrows in Figure 2e, reflecting their opposite chirality from the broken mirror symmetry.

**Electronic properties and circular photo-galvanic effect of Te/graphene devices.** We first fabricated devices of a Te nanowire connected by graphene electrodes by polymer-assisted graphene transfer, electron beam lithography, and metal deposition (see Methods for details). Figure 3a shows a 3D AFM image of a typical device of a Te nanowire with graphene electrodes. The length of the nanowire is 45 µm, and the height is around 150 nm as shown in the line profile. Figure 3b shows the output characteristics of a Te/graphene device exhibiting linear I-V relation. Our Te/graphene devices on 90 nm $SiO_2$/Si substrates serving as a back-gate exhibited p-type semiconducting behaviors as shown in Figure 3c. We also measured the circular photo-galvanic effect (CPGE) of the devices to demonstrate the chirality-dependent characteristics of the devices [31,32]. The measurement scheme is shown in Figure 3d. Photocurrent along a chiral Te nanowire at zero bias voltage was measured with varying quarter-wave plate angle λ under obliquely incident light (θ = 45°) at the center of the nanowire (Figure 3e). Here, λ = 45° and λ = 135° represent right circular-polarization (RCP) and left circular-polarization (LCP) of the incident light, respectively. The polarization-dependent photoresponse as a function of quarter wave-plate angle λ was fitted with the below equation following previous studies[31,33].

$$I = C \sin 2\lambda + L_1 \sin 4\lambda + L_2 \cos 4\lambda + D$$

The coefficients of C, $L_{1,2}$, D refer to a circular polarization dependent photocurrent, a linear polarization dependent one, and a polarization independent one, respectively. Circularly polarized light selectively excites electrons of specific spin polarization in a Te nanowire coupled to momentum, attributed to spin-dependent band-splitting in a Te nanowire as shown in Figure 2e[31,32]. Thus, the spin-momentum coupling from the chirality of Te nanowires gives rise to different photocurrent dependent on the circular polarization of incident light along the Te nanowire, i.e. a longitudinal CPGE. The chirality of the Te nanowires determined by the CPGE is consistent with the one confirmed by STEM images of the cross-section of the nanowires[28,42,43].

**Reflective circular dichroism measurements.** To investigate CISS in chiral Te nanowires, RMCD is measured with charge current along the nanowires. CD is different absorption of left-circularly polarized (LCP) light and right-circularly polarized (RCP) light in a material, indicative of broken mirror symmetry (i.e., natural CD) or broken spin degeneracy due to magnetic field (i.e., magnetic circular dichroism, MCD)[44]. CD has been used to determine the chirality of biological materials and synthetic meta-materials as well as magnetization of materials[45,46] and spin accumulations[47]. The observed RMCD in previous studies[45-47] is attributed to the interplay between the spin angular momentum of light and the asymmetric spin population of materials, originating from the inversion or the time reversal symmetry breaking in the system. Thus, RMCD is a useful tool to study localized spin polarization at the location of incident light, allowing real-space imaging of spin polarization by scanning the incident light. It is also employed in our work to observe spin polarization in chiral Te nanowires induced by charge current, *i.e.* CISS.

To measure the spin polarization along the atomic chain axis of a Te nanowire, the incident laser was tilted at an angle of θ = 30° with aligning the nanowire in the plane of incidence. Then, the difference of the reflection of left- and right-circularly polarized light from the nanowire ($R_{LCP}$ - $R_{RCP}$) was measured while applying DC current along Te nanowires with different chirality (Figure 4a). Figure 4b shows RMCD as a function of applied current. Here, the RMCD at $I = 0$ was subtracted to offset. RMCD changes linearly with applied current, indicating spin polarization in the Te nanowire was linearly changed with charge current. In addition, the sign of RMCD reversed as the direction of charge current reversed, showing spin polarization was flipped with the reversed charge current. It should be noted RMCD was not changed with the change of current when light was incident in the normal direction, i.e. θ = 0° (Figure S1), confirming the direction of the polarized spin is along the nanowires. Importantly, the slope of RMCD depending on charge current was opposite for devices with opposite chirality (Figure 4b), which confirms the formation of opposite spin polarization on the opposite chirality of Te nanowires. The spin polarization dependence on charge current and chirality characterizes the CISS effect in Te nanowires[28,37-39].

**Real-time time-dependent DFT calculations.** The spin dynamics induced by current and the origin of CISS are also studied through the ab initio methods using real-time time-dependent DFT (rt-TDDFT). The calculation methods are described in the Methods section. The charge-neutral band structure itself has a marginal band gap, as shown in Figure 2e, and we adjusted the Fermi level by assuming a doping of 0.1 hole per unit cell, which corresponds to the hole doping condition of Te nanowires (Figure S2). We applied an external electric field along the chiral axis and calculated the time profile of the charge current and spin polarization over time obtained from the time-evolving Kohn-Sham states of the rt-TDDFT. Figure 4c depicts the applied external electric field and the resulting current and spin. The electric field is applied in the axial direction of the wire, and we tested with two opposite directions of electric field together with two opposite handedness of the wire. As shown in Figure 4d, the current develops with the external electric field, but the resulting spins of a given current are opposite for opposite handedness, consistent with experimental results. With separate calculations without SOC, zero spin was observed in the wire. Thus, the SOC and chiral geometry are indispensable factors for CISS in Te nanowires.

**Real space imaging of spin polarization in a device of a Te nanowire and graphene electrodes.** As already noted, a spin scattering mechanism governing CISS can be confirmed by observing spin configuration at the interface between chiral materials and electrodes in real space (Figure 1). To experimentally observe the spin configuration, we performed scanning RMCD on a device shown in Figure 5a. Herein, it should be noted that we used graphene with negligible SOC instead of metal as electrodes in order to study spin scattering at the interface without interference from electrodes[48] because conventional metal materials used as electrodes interfere the spin configuration at the chiral material/electrode interface due to strong SOC. Furthermore, the long spin lifetime in graphene due to weak SOC provides an advantage for optical detection. We mapped RMCD to obtain real space spin polarization over the whole system of a Te nanowire and graphene electrodes with different current (Figure 5b and 5c). Of interest, non-zero RMCD was observed not only on the Te nanowire but also around the interface between the nanowire and graphene. The RMCD was reversed when the direction of the current along the nanowire was reversed. More importantly, the spin polarization of the nanowire and that of graphene electrodes are same for both cases, which is strongly indicative of the spin polarizer mechanism in CISS in the system of a Te nanowire and graphene electrodes. We performed the RMCD scanning experiment on a different device with a Te of opposite chirality (Figure 5d, 5e, and 5f). In this case, we also observed the same polarization on the chiral nanowire and graphene electrodes, however, the spin polarization for given polarity of applied current was opposite due to the opposite chirality.

In addition, the observed RMCD is strongest at the interface and decreases away from the interface, indicating spin relaxation in the graphene. The measured RMCD is analyzed as a function of distance from the interface fitting to exponential decay with the characteristic spin relaxation length of 7.74 μm and 6.34 μm in case of applying positive and negative current along the wire, respectively (Figure 5g and 5h). The characteristic spin relaxation length is in good agreement with the ones of multilayered graphene on $SiO_2$ (~3~8 μm) in previous works measured by nonlocal spin detection method in spin valve systems[49,50]. The spin relaxation length for the device of opposite chirality was measured as 3.74 μm and 5.37 μm, which is also in a similar range, as shown in Figure S3.

DISCUSSION

Our observation resolves the debate between spin filter and spin polarizer regarding CISS. Injecting electrons from graphene are non-spin polarized and get scattered by a Te nanowire. The spin filter scenario presumes that two spin channels are independent so that one channel prefers to transmit while the other channel is reflected. However, two spin channels should mix (spin flips) in the presence of SOC of a Te nanowire, which does not support the assumption of spin filter. In fact, both transmitted and reflected electrons are strongly scattered by a Te nanowire. If one spin channel favors transmission after scattering, the opposite spin channel favors reflection but requires a spin flip in the backscattering due to SOC (Figure 5i), as predicted by recent calculations[19,20]. This spin-polarizer mechanism in CISS is supported by the same spin-polarization at two graphene electrodes in our observation.

Notably, different from previous studies that demonstrated the magnetization in Te due to Rashba-Edelstein effect[28], our work focuses on the CISS-driven spin polarization for incoming and scattering electrons in two electrodes rather than Te itself. CISS refers to the induced-spin polarization at source and drain electrodes, outside the Te nanowire, as illustrated by Figure 1. In other words, we revealed the parallel relation between two graphene electrodes rather than the magnetoresistance or spin polarization inside the chiral material itself due to the Edelstein effect. The spin distribution on the interface is the key observation to reveal the CISS mechanism.

It should be also noted that our system of a Te nanowire with strong SOC along with graphene electrodes with negligible SOC is a promising platform to study CISS mechanism. We also note the validity of our material system for CISS. Most CISS experiments were conducted on organic chiral molecules. Because of negligible intrinsic SOC in these chiral molecules, it is impossible to rationalize the high spin polarization ratio (commonly 60% or even higher) claimed in experiments. Therefore, recent efforts have been devoted to designing enhanced SOC[10], for example, by substrate proximity[25-27], electron-phonon interaction[51], nuclei-vibration beyond Born–Oppenheimer (BO) approximation[52], or geometric SOC etc[53]. Overall, all these theories aim to reveal strong SOC in chiral molecules to explain the CISS effect, which has been hidden in conventional treatments such as BO-approximation, and realize an effective spin filter. To avoid the mysterious origin of SOC, we choose chiral Te nanowires with intrinsically strong SOC, which is equivalent to chiral molecules with some effective SOC. Therefore, the Te nanowire with undebated strong SOC is a promising chiral system to investigate the spin polarization caused by CISS, beyond searching for SOC origin in chiral molecules.

CONCLUSION

In conclusion, we reveal the spin scattering mechanism, the key concept in CISS, for chiral Te nanowires through real-space imaging of spin distribution on the nanowires and the interface with graphene electrodes using magneto-optical technique. The in-plane assembly of chiral materials and the graphene electrodes of negligible SOC enables real space imaging of the entire system – chiral nanowire and graphene electrodes – without interference from the electrodes. Our observations and time-dependent first-principles calculations evidence that the CISS in Te nanowires are attributable to the interplay of the chirality and strong SOC. Of note, we also demonstrate the current-dependent spin polarization around the interface between Te nanowires and graphene electrodes, where the spins in the nanowire and the electrodes share the same polarity. This is strongly indicative of the spin polarizer mechanism, as opposed to the spin filter one, in CISS. Our direct observation of spin distributions at chiral Te nanowires and graphene electrodes with magneto-optic method opens the path to understand the spin polarization mechanism of CISS and exploit it to chirality-based optoelectronic and spintronic devices.

METHODS

**Chemical synthesis of Te nanowires.** 105 mg of $Na_2TeO_3$ and 548 mg of polyvinylpyrrolidone (average molecular weight, 360,000) were dissolved in 33 ml of MilliQ water. 3.65 mL of $NH_4OH$ solution (25% by weight in water) and 1.94 mL of hydrazine hydrate (80% w/w) were added while stirring. The mixture was sealed in an autoclave and heated at 180 °C for 23 h. The resulting solution was precipitated by centrifugation (4,000 r.p.m., 5 min) and washed three times with distilled water.

**Device fabrication.** Te nanowires were transferred onto a 90 mm $SiO_2$/Si substrate using a simple drop-casting method. The two-terminal devices were patterned using electron beam lithography, few-layer graphene flakes were exfoliated and transferred to the ends of the Te nanowires, and 5/50 nm Cr/Au metal contacts were used for electric connection to the graphene electrodes.

**Photocurrent measurements.** A 532 nm laser beam was used with an objective to a spot size of ~2 μm and each incident angles. The laser spot was scanned along the whole device using a galvo mirror system. A liquid crystal variable retarder was used to control the quarter wave-plate angle (λ) of the beam to change the circular polarization. The photocurrent was measured using a source meter (Keithley 2450). All measurements were performed at zero source-drain bias and room temperature.

**Reflective magnetic circular dichroism measurement.** The reflective magnetic circular dichroism measurement was performed in a cryostat at 3.5 K. The DC current was applied to the device using a source meter (Keithley 2450). A 532 nm laser beam with a spot size of ~2 μm was used to probe the sample at obliquely incident angle of 30°. The laser spot was scanned along the whole device using a galvo mirror system. A photoelastic modulator (PEM) was used to control the polarization angle of the beam at 50 kHz. The amplitude difference between right-circularly polarized light and left-circularly polarized light of reflected beam data was acquired using a lock-in amplifier (Stanford Research Systems, SR830).

**Structural characterization.** Two devices with opposite chirality-dependent photocurrent and slope of reflective magnetic circular dichroism were selected for STEM chirality characterization. Te nanowires lamellae were prepared using Ga+ ion beam milling with a dual-beam focused ion beam (FIB) system (Thermo Fisher Scientific, Helios Nano Lab 450). The HAADF-STEM images were acquired using a Cs-corrected Titan Cube G2 60–300 (Thermo Fisher Scientific) operating at 200 kV. For the determination of chirality, STEM images were collected with the sample aligned along the [010] zone axis and then rotated by 30° for mirrored images. The handedness was determined by comparing with known chiral atomic models.

**Calculation on band structure and spin behavior in chiral Te.** We performed the static density functional theory (DFT) calculation using the Quantum ESPRESSO package[54,55] to obtain the self-consistently converged electronic ground state of the left-handed and right-handed 3 Te chiral chains. For the exchange-correlation energy of the system of electrons, we used Perdew-Burke-Ernzerhof (PBE)-type generalized gradient approximation (GGA)[56]. The fully relativistic norm-conserving pseudopotential is employed to describe the potential of ions and the spin-orbit coupling effect. The energy cutoff for the plane-wave-basis expansion was set at 60 Ry. The Brillouin zone is sampled through the Monkhorst-Pack scheme with a 9×9×11 grid. We also performed real-time time-dependent DFT (rt-TDDFT) calculations using our in-house code to examine the electric current and magnetization time profile[57-59]. In the time evolution, we self-consistently evolved the time-dependent Kohn-Sham wavefunctions, density, and Hamiltonian with the time-dependent Kohn-Sham equation as below.

$$i\hbar \frac{\partial}{\partial t}\psi_{n,k}(r,t) = \left[\frac{1}{2m}\left(-i\hbar \nabla + \frac{e}{c}A_{ext}(t)\right)^2 + \sum_{\lambda} V_{atom}(r - R_\lambda) + V_{DFT}[\rho(r,t)]\right]\psi_{n,k}(r,t)$$

Where $n$ and $k$ denote the band index and the Bloch momentum vector, respectively. $A_{ext}(t)$ and $V_{DFT}$ indicates the external time-dependent vector potential and DFT potential which contains the exchange correlational potential that is the functional of the time-dependent density. The uniform electric field is represented in the velocity gauge as a vector potential that was previously mentioned through the relation $E = -\frac{1}{c}\frac{\partial}{\partial t}A(t)$. The discretized time step ($\Delta t$) is set to 2.415 ato-seconds. The time profile of current and spin are obtained from the time-evolving Bloch states as follows:

$$I_z(t) = -\frac{e}{m_e}\sum_{n,k} f_{n,k}\langle \psi_{n,k}(t)|\hat{\pi}_z|\psi_{n,k}(t)\rangle, S_z(t) = \sum_{n,k} f_{n,k}\langle \psi_{n,k}(t)|\hat{S}_z|\psi_{n,k}(t)\rangle$$

where the $f_{n,k}$ is occupation of the ground Bloch state which obtained from the DFT calculation, and $m_e$ is the mass of electron. For the calculation of current, we used the gauge-invariant mechanical momentum $\hat{\pi} = \frac{m_e}{i\hbar}[\hat{r},\hat{H}] = \hat{p} + \frac{e}{c}A_{ext}(t) + i\frac{m_e}{\hbar}[\hat{V}_{NL},\hat{r}]$[59]. To treat the light-induced field correction to the relativistic effect, we considered the correction term $[\sim e\sigma \cdot (A(t)\times \nabla V)]$ and substrate it from the relativistic effect of DFT.

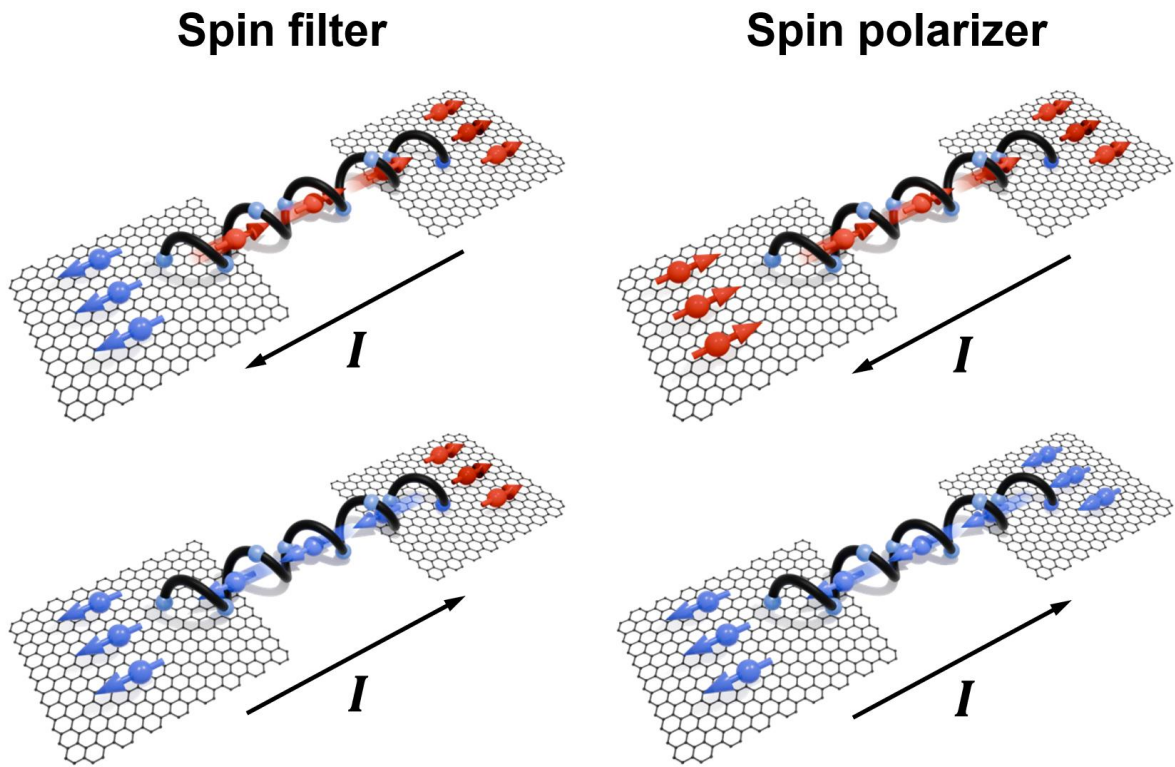

**Figure 1.** Plausible candidate mechanisms to explain chirality-induced spin selectivity (CISS) in non-magnetic chiral materials. Spin filter (left) and spin polarizer (right) mechanism to explain the change of spin polarization of the chiral materials at center depending on the different direction of charge current. Note the spin polarization at both electrodes is opposite on both electrodes (i.e. spin current along the chiral material) for spin filter model, while it is same for spin polarizer model.

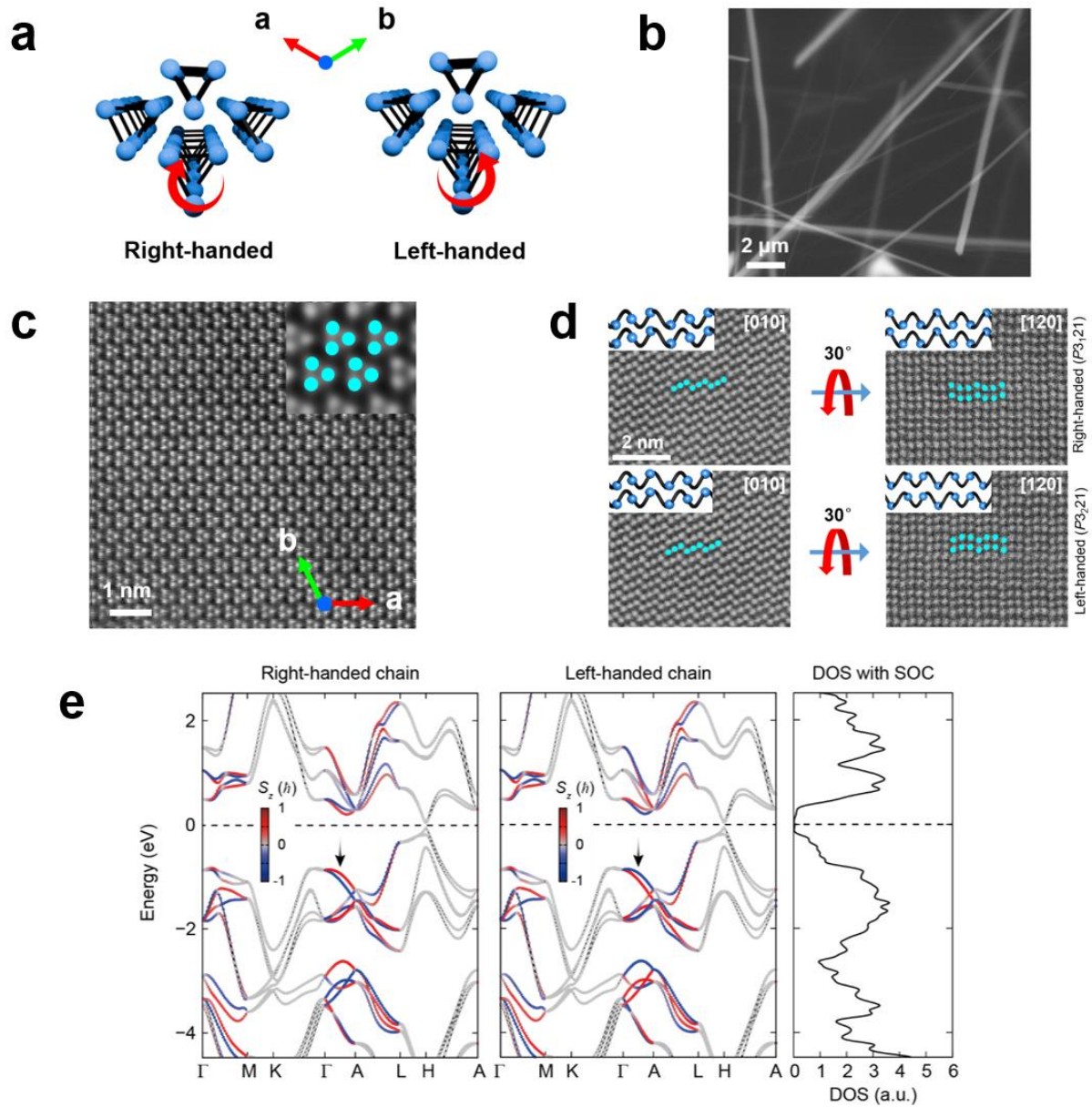

**Figure 2.** Chiral crystal structure of tellurium (Te) nanowires. (a) The crystal structure of trigonal Te. (b) Scanning electron microscopy image of Te nanowires drop-casted on a SiO$_2$/Si substrate. (c) STEM image of a Te nanowire cross-section normal to the c axis. (d) STEM images of two Te nanowires with opposite chirality, showing [010] and [120] zones of the right-handed (top) and left-handed (bottom) Te nanowires. The chirality is confirmed by the rotation of 30º. Inset, schematic crystal structures of two Te NWs with opposite chirality. (e) Spin-resolved electronic

band structures of right-handed and left-handed trigonal Te nanowires, calculated with spin–orbit coupling. The color scale denotes the out-of-plane spin component $S_z$. The energy window is referenced to the Fermi level (dashed line). The rightmost panel shows the corresponding total density of states (DOS)

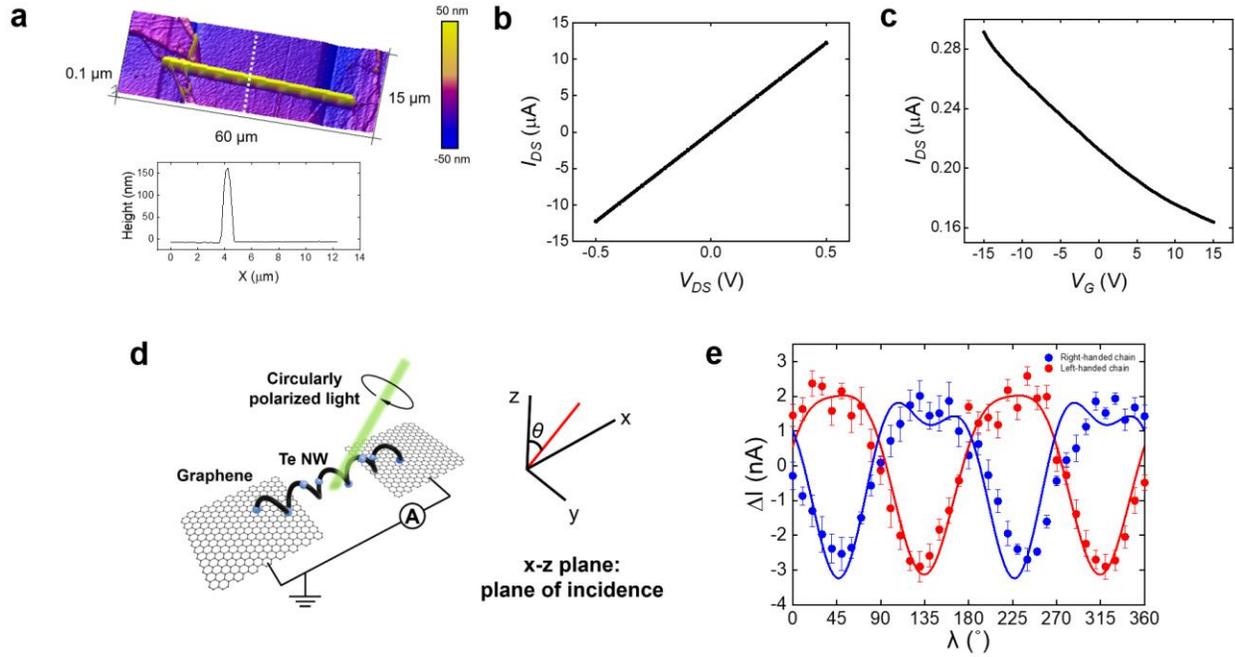

**Figure 3.** Electronic properties and circular photo-galvanic effect of Te/graphene devices. (a) 3D AFM image of a Te/graphene device. Output characteristic (b) and transfer characteristic (c) of a Te/graphene device. (d) Schematic image of a setup to measure circular photo-galvanic effect of a Te/graphene device. (e) Chirality-dependent photocurrent as a function of a quarter-wave plate angle λ of right- and left-handed Te nanowires under obliquely incident light (θ = 45°). An opposite circular photo-galvanic effect is induced by the opposite handedness of the Te nanowire.

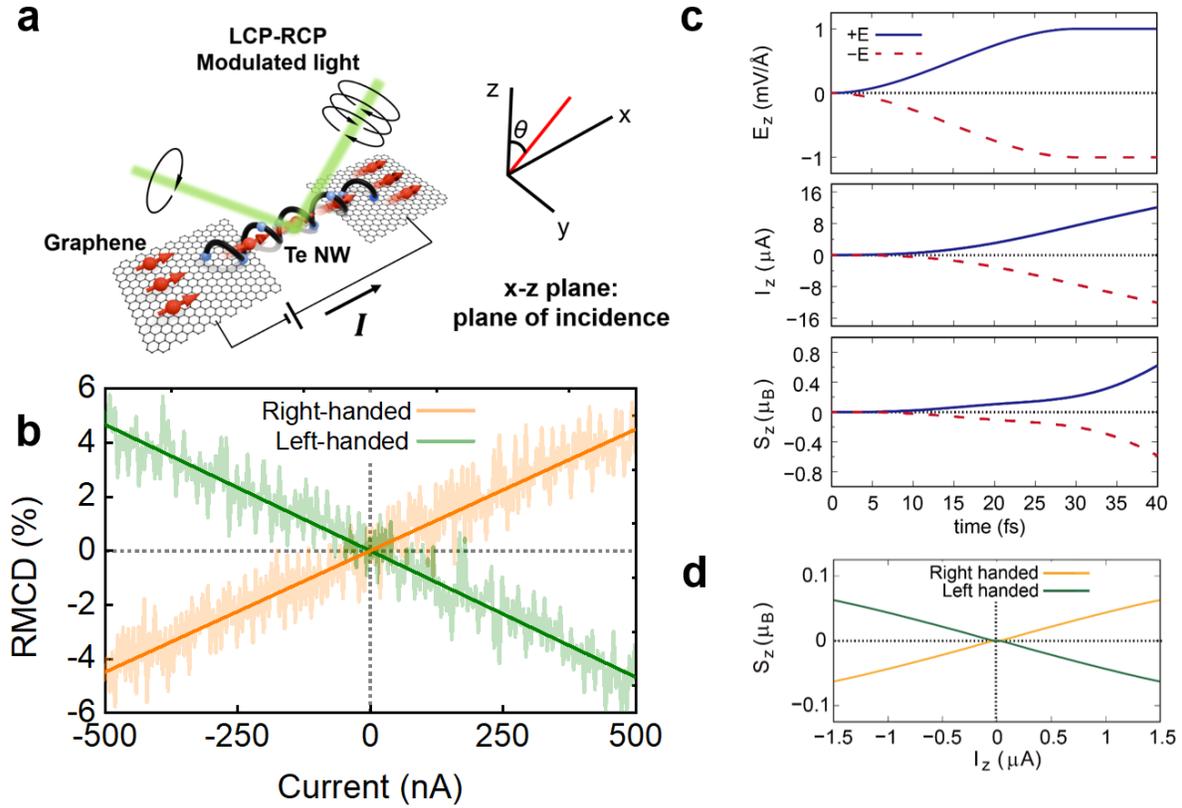

**Figure 4.** CISS in Te nanowires. (a) Schematic of experimental setup for RMCD measurement. (b) Normalized RMCD intensity as a function of applied DC current along right- and left-handed Te nanowires. Note the opposite slope for opposite chirality. (c) Time-dependent spin per atom ($S_z$) and charge current ($I_z$) under an applied external electric field calculated by rt-TDDFT (d) Spin per atom ($S_z$) with respect to charge current ($I_z$). The same calculations were performed with nanowires with opposite handedness, and we only present those of right-handed one in (c).

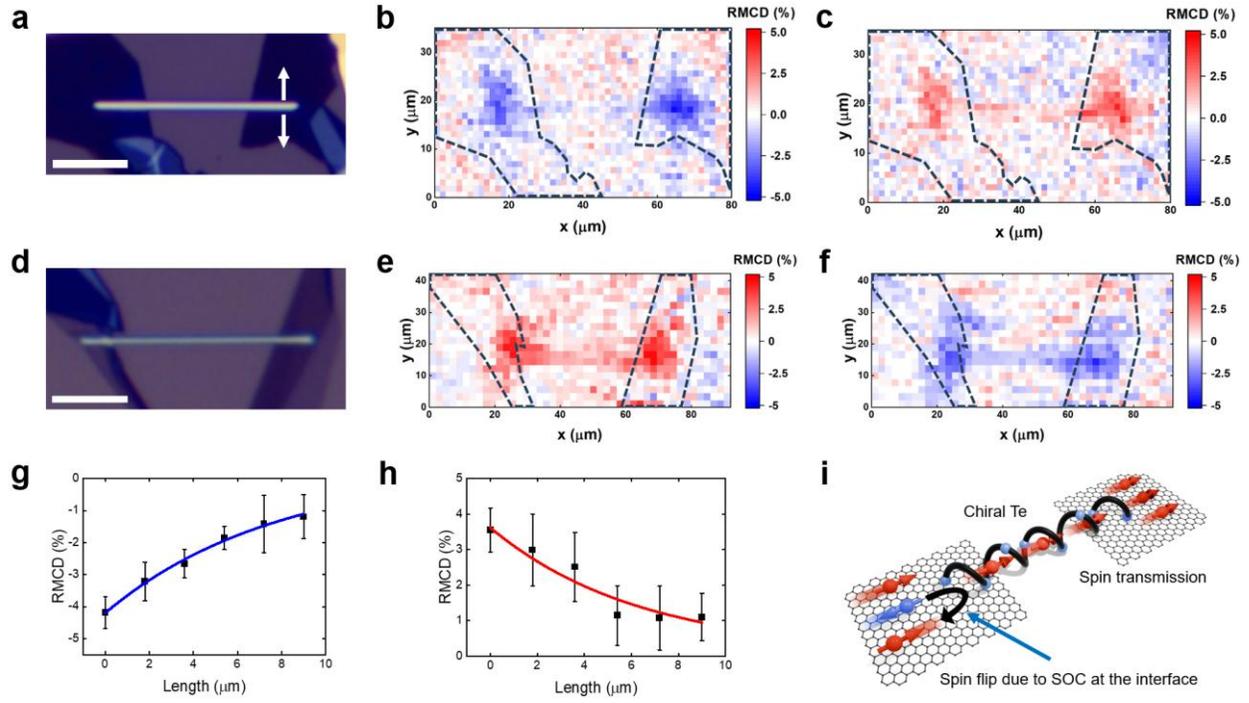

**Figure 5.** Real space imaging of spin polarization in a device of a Te nanowire and graphene electrodes. (a) Optical image of a device of Te nanowire/graphene electrodes. The white scale bars correspond to 20 μm. RMCD maps for the device (a) with applied current of +750 nA (b) and -750 nA (c). (d) Optical image of a device of Te nanowire/graphene electrode with opposite chirality. RMCD maps for the device (d) with applied current of +500 nA (e) and -500 nA (f). (g), (h) RMCD as a function of distance from the wire from (b) and (c), respectively. Mean values from eight line profiles (along the white arrows in (a)) are shown, and s.d. values are shown with error bars. The data is fitted to an exponential function, and the characteristic length (i.e. spin relaxation length for both cases) is 7.74 μm and 6.34 μm for (g) and (h), respectively. (i) Spin scattering mechanism at the interface of a Te nanowire/graphene electrode.

ASSOCIATED CONTENT

**Supporting Information**.

The following files are available free of charge.

Figure S1: Normalized reflective magnetic circular dichroism (RMCD) intensity with different incident angles; Figure S2: The electronic band structure of the bulk Te; Figure S3: RMCD as a function of distance for the device of opposite chirality.

AUTHOR INFORMATION

**Corresponding Author**

*Seon Namgung, *Noejung Park, *Binghai Yan

**Author Contributions**

S.N. supervised the project. J.L. synthesized the Te crystals and prepared the devices with the help of D.D. and performed the reflective magnetic circular dichroism measurement with the help of S.L. and analyzed data under the guidance of S.N. J.L. performed the photocurrent measurement and analyzed data. J.L. and D.D. measured AFM data of the devices. U.J. performed the density functional theory calculations under the guidance of N.P. and B.Y. S.S. performed the scanning transmission electron microscopy under the guidance of H.Y.J. S.N., N.P., H.J., and B.Y. discussed the results and developed the explanations on the experiments. The manuscript was written through contributions of all authors. All authors have given approval to the final version of the manuscript. ‡These authors contributed equally.

ACKNOWLEDGMENT

This work was supported by the National Research Foundation of Korea (NRF) grant funded by the Korea government (MSIT) (No. RS-2023-00218799 and No. RS-2022-NR069722). We thank UNIST Central Research Facilities (UCRF) for the support of its facilities and equipment. We also thank Prof. Dai-Sik Kim (UNIST) for the support of equipment.

Supplementary information for

# Real Space Imaging of Spin Scattering in Chirality-Induced Spin Selectivity

*Jaehyun Lee[1]‡, Sang-hyuk Lee[1]‡, Uiseok Jeong[1], Daryll J. C. Dalayoan[1], Soobeom Shin[2], Hu Young Jeong[2], Hosub Jin[1], Binghai Yan[3,4]\*, Noejung Park[1, 5]\*, Seon Namgung[1]\**

[1]Department of Physics, Ulsan National Institute of Science and Technology (UNIST), Ulsan, Republic of Korea

[2]Graduate School of Semiconductor Materials and Devices Engineering, Ulsan National Institute of Science and Technology (UNIST), Ulsan, Republic of Korea

[3]Department of Condensed Matter Physics, Weizmann Institute of Science, Rehovot, Israel

[4]Department of Physics, Pennsylvania State University, PA, US

[5]Max Planck Institute for the Structure and Dynamics of Matter, Hamburg, Germany

*Correspondence to: seon@unist.ac.kr, noejung@unist.ac.kr, binghai.yan@weizmann.ac.il

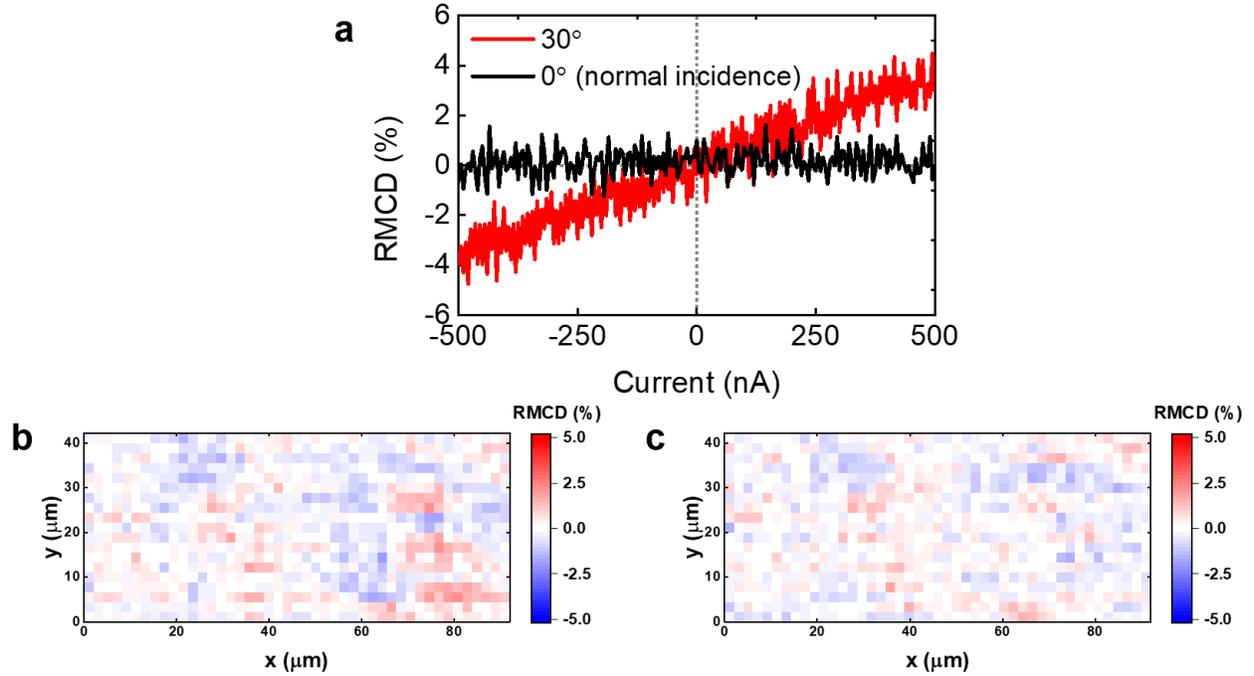

**Figure S1.** (a) Normalized reflective magnetic circular dichroism (RMCD) intensity as a function of the applied DC current in a right-handed Te nanowire with different incident angles. RMCD mapping data in the Te/GR device at normal incidence of light with applied current of +500 nA (b) and -500 nA (c) RMCD was not changed with current change at normal incidence of light, indicating the direction of the polarized spin is along the nanowires.

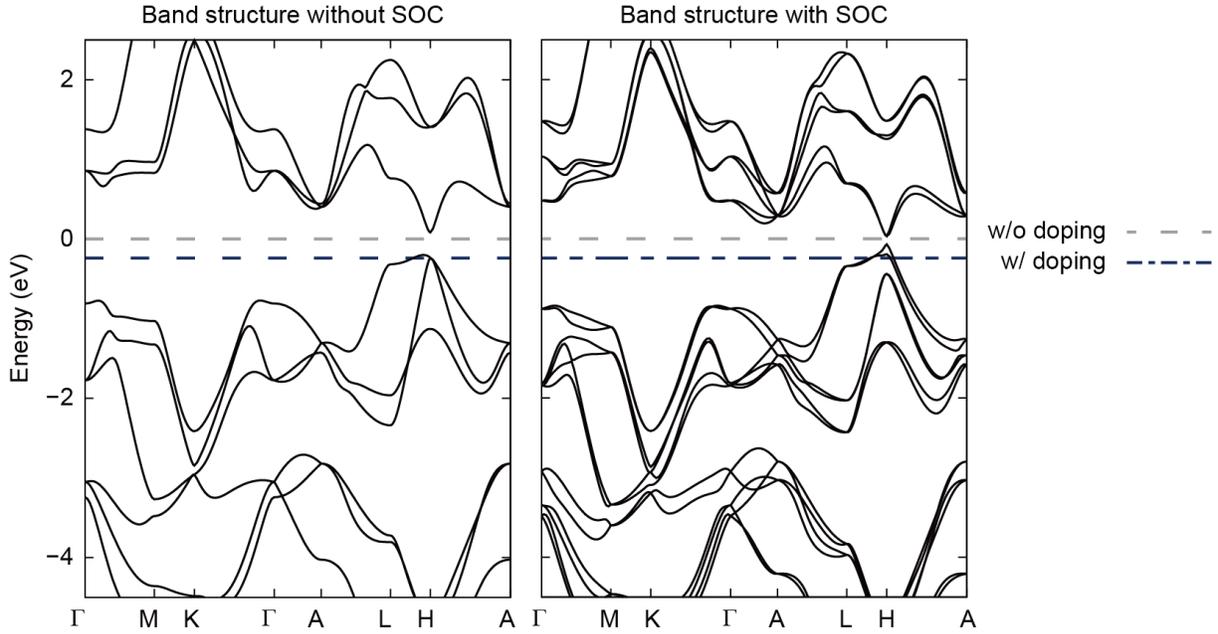

**Figure S2.** Electronic band structures of Te wire calculated using static DFT, as described in the Methods section. The left panel shows the band structure without SOC, while the right panel includes SOC, where the characteristic band splitting emerges. To simulate a metallic state and capture the current-induced spin dynamics, a hole doping of 0.1 per unit cell was introduced. The grey dashed line indicates the Fermi level of the charge-neutral system, whereas the dark blue dashed line corresponds to the Fermi level under hole doping.

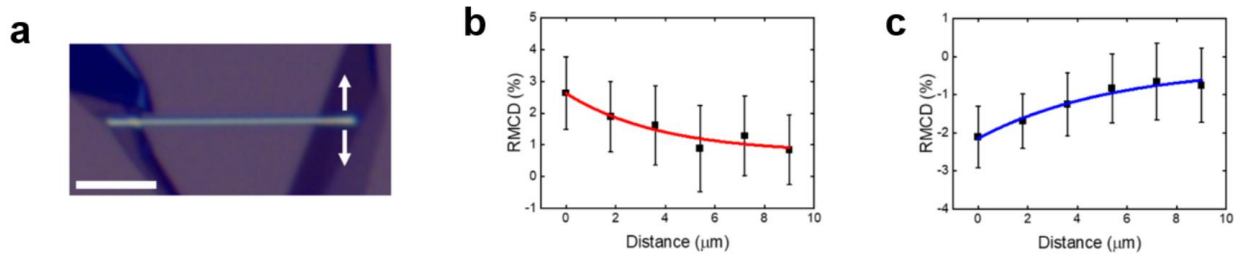

**Figure S3.** RMCD as a function of distance from the nanowire with applied current of +500 nA (b) and -500 nA (c) from the device of Te nanowire/graphene electrodes shown in (a). Mean values from eight line profiles (along the white arrow in (a)) are shown, and s.d. values are shown with error bars. The data is fitted to an exponential function, and the characteristic length (i.e. spin relaxation length for both cases) is 3.74 μm and 5.37 μm for (b) and (c), respectively.